\documentclass[aps,twocolumn,preprintnumbers,amsmath,amssymb,floatfix,groupedaddress,nofootinbib]{revtex4}

\usepackage{enumerate}
\usepackage{graphicx}
\usepackage{color}

\usepackage[applemac]{inputenc} 
\usepackage{fontenc}  
\usepackage{amsmath}

\newcommand{\beq}{\begin{equation}}
\newcommand{\eeq}{\end{equation}}
\newcommand{\bea}{\begin{eqnarray}}
\newcommand{\eea}{\end{eqnarray}}
\newcommand{\ba}{\begin{array}}
\newcommand{\ea}{\end{array}}

\newcommand{\bef}{\begin{figure}}
\newcommand{\eef}{\end{figure}}

\begin{document}

\title{The Einstein-Bohr debate: finding a common ground of understanding~? }

\author{Nayla Farouki$^{(1)}$ and Philippe Grangier$^{(2)}$}

\affiliation{ 
(1): Philosopher and historian of science, http://www.homo-rationalis.com, France. \\
(2): Laboratoire Charles Fabry, IOGS, CNRS, Universit\'e Paris~Saclay, F91127 Palaiseau, France.}

\begin{abstract}
After reminding the main issues at stake in the famous Einstein-Bohr debate initiated in 1935, we tentatively propose a  way to get them closer, thus shedding a new light on this historical discussion.

\end{abstract}

\maketitle

\subsection{The historical debate.} 

Realism in physics is often understood as the idea that all physical objects are built up by assembling smaller (or more elementary) objects, and that all objects carry their own properties, whatever their level in this construction. This idea works fine in classical physics, but is clearly more problematic in quantum physics, where the nature of objects becomes more elusive, and where it is dubious that a single object carries its own properties, especially when it is considered to be part of an entangled set. 

Unhappy with this elusive character of quantum mechanics (QM), Einstein, Podolsky and Rosen (EPR) expressed their own view of realism \cite{EPR} in a stronger and more definite statement, saying that, ``If, without in any way disturbing a system, we can predict with certainty (i.e., with probability equal to unity) the value of a physical quantity, then there exists an element of physical reality corresponding to this physical quantity''. 

Yet, Niels Bohr disagreed. In his answer, written also in 1935  \cite{Bohr1935}, he wrote: ``EPR's criterion of physical reality contains an ambiguity in the meaning of ‘without in any way disturbing a system'. Of course there is no mechanical disturbance of the system under investigation, but there is an influence on the {\bf very conditions} which define the possible types of predictions regarding the future behavior of the system. These conditions constitute an inherent element of the description of any phenomenon to which the term ‘physical reality’ can be properly attached''. These statements by Bohr were mostly accepted, but not really understood: what are these ``very conditions'' quoted as inherent elements to speak about physical reality? We will try to clarify this issue below. 

In the 1960’s, Bohr gave his own full-fledged definition of what physics should be \cite{Bohr1963}: ``Physics is to be regarded not so much as the study of something a priori given, but rather as the development of methods of ordering and surveying human experience. In this respect our task must be to account for such experience in a manner independent of individual subjective judgement and therefore objective in the sense that it can be unambiguously communicated in ordinary human language''. But this latter statement sounded extreme, and renouncing the very idea of a physical object is a hard price to pay for most physicists – this is even more the case nowadays, given the impressive progress achieved in getting full control of individual quantum objects.

In 1964, an important advance took place \cite{Bell}. Theoretical and experimental work, initiated by John Bell, showed that the EPR view could not work as such, and must at least be amended \cite{Aspect}. Quantum reality does not abide – at least not entirely – by Einstein’s conditions. As Bohr’s arguments were either hard to accept or equivocal, the debate went back to its initial state, and new questions and theories about the ``real objects'' involved in a quantum description kept showing up \cite{Laloe}.

\subsection{Physical realism and contextual objectivity.} 

Yet, if absolute certainty and total predictability could not be the conditions required to attain quantum reality, perhaps, a different set of rules could make this reality convincing again? This is what is proposed here. First, we define physical realism with the more consensual statement that {\bf ``the purpose of physics is to study entities of the natural world, existing independently from any particular observer’s perception, and obeying universal and intelligible rules''.} Universality and intelligibility are accepted criteria for scientific realism, perhaps not as strong as full certainty and predictability, but they may be more efficient in the QM realm. This being done, would it be possible to generalize the argument for universality and intelligibility so as to admit at least parts of Einstein’s and Bohr’s positions? In an attempt to define what is real in QM, and to get Einstein and Bohr closer, let us first take into account the evidence that can be gathered from experience, while provisionally putting aside the mathematical apparatus through which this experience is described.

The answer we propose \cite{csm1} relies on the simple observation that in quantum physics, like in classical physics, the values of some physical properties can be measured repeatedly, and the results can be predicted with certainty. But in quantum physics, this can generally be obtained only by keeping constant all the relevant elements of the measurement procedure, that we call a context. If we admit that this complete classical specification of the context is just the ``very condition'' required by Bohr, then the following statement, paraphrasing Einstein’s, is perfectly fine, also from a quantum point of view:  ``If, without in any way disturbing a system, {\bf nor changing the context}, we can predict with certainty (i.e., with probability equal to unity) the value of a physical quantity, then there exists an element of physical reality corresponding to this physical quantity''.

This statement satisfies EPR’s idea to look for certainty in scientific knowledge, as a criterion for reality \cite{csmBell}. The ``element of physical reality'' that Einstein sought is present in a deterministic way, but it is submitted to a condition not perceived in classical physics: the statement is {\bf true as long as the context is not changed} \cite{note}. So the element of quantum reality belongs jointly to the (quantum) system and to the (classical) context, and is real as EPR would have wanted, but its reality comes with a twist: it permanently associates a quantum system and a classical context. To distinguish it from other forms of classical realities, we shall call it a modality.

When associated with quantization – postulated as an intrinsic property of quantum objects \cite{csm1} – a change of context will produce a random change affecting the modalities of a quantum object, when it shows up in a different context \cite{csm1,csmBell}. This randomness appears in the form of probabilities, and it can be shown that they can be computed according to the usual quantum formalism \cite{csm2,csm3}. In this approach, the modality is a real phenomenon involving a system and a context as physical objects, and it is thus ontologically distinct from the state vector or projector, which is a mathematical object used to calculate probabilities. Mixing up these two notions – the real phenomenon on the one hand and the mathematical object used to describe it on the other – has been generating a lot of confusion, which is removed within the quantum realism framework proposed here. 

\subsection{What did we learn ?} 

Our contextual description also tells something about Einstein’s and Bohr’s respective positions. It does agree with Einstein’s realism in terms of certainty, necessity, and therefore physical reality. At the opposite, it clashes with his classical conviction that physical properties should be independent of the context in which they are defined. We have also a good agreement with Bohr’s intuition that the conditions of the measurement must be taken into consideration: some properties do vary randomly when the context is changed. We attribute this random change to the fact that elementary physical  properties are quantized, implying that randomness is indeed essential to understand quantum physical behavior. This randomness, famously, was refused by Einstein. Though a final reconciliation between Einstein and Bohr seems out of scope, both Einstein’s objectivity and Bohr’s contextuality do find a common ground in our approach. 

Does this new understanding answer the major challenge of defining quantum reality? We believe that this is the case, since certainty exists, quantum probabilities make sense due to quantization, and Bohr’s intuition that we ought to consider the experimental conditions as an inherent part of the object under study is given its full meaning. This only requires that we describe objectivity in a new more humble way: there is no such a thing as absolute objectivity, i.e. the apprehension of the Universe, taken as a whole, with all its logical and ontological necessity as Newton and Einstein conceived it. Contextual objectivity \cite{CO2002} may provide us with true and efficient knowledge about the world we live in.

\vskip 2mm

{\bf Acknowledgements.} 
The authors thank Alexia Auff\`eves for continuous contributions, and Franck Lalo\"e \& Roger Balian for many useful discussions.


\vskip 2mm

\begin{references}


\bibitem{EPR} A. Einstein, B. Podolsky, and N. Rosen, ``Can Quantum-Mechanical Description of Physical Reality Be Considered Complete?'', Phys. Rev. 47, 777 (1935).

\bibitem{Bohr1935} N. Bohr, ``Can Quantum-Mechanical Description of Physical Reality be Considered Complete?'', Phys. Rev. 48, 696 (1935).

\bibitem{Bohr1963}  N. Bohr, ``The Unity of Human Knowledge'', in ``Essays 1958-1962 on Atomic Physics and Human Knowledge'' (Ox Bow Press, 1963)
\bibitem{Bell}  J.S. Bell,  ``On the Einstein-Podolski-Rosen paradox'', Physics 1, 195 (1964).

\bibitem{Aspect}  A. Aspect,  ''Closing the Door on Einstein and Bohr's Quantum Debate'', Physics 8, 123 (2015)

\bibitem{Laloe}  F. Laloë,   ``Do We Really Understand Quantum Mechanics?'', Cambridge University Press (2012).

\bibitem{csm1} A. Auff\`eves and P. Grangier, ``Contexts, Systems and Modalities: a new ontology for quantum mechanics'', Found. Phys. 46, 121 (2016) [arXiv:1409.2120].

\bibitem{csmBell}  A. Auff\`eves and P. Grangier,  ``Violation of Bell's inequalities in a quantum realistic framework'', Int. J. Quantum Inform. 14, 1640002 (2016) [arXiv:1601.03966].

\bibitem{note} Strangely enough, this experimental fact has been known all along, but the idea that the system's properties should be defined independently from the context is so well entrenched that this plain fact has constantly been ignored. As we show here, its mere existence is decisive.


\bibitem{csm2} A. Auff\`eves and P. Grangier,  ``Recovering the quantum formalism from physically realist axioms'', Scientific Reports 7, 43365 (2017) [arXiv:1610.06164].

\bibitem{csm3}  A. Auff\`eves and P. Grangier,  ``Extracontextuality and extravalence in quantum mechanics",  Phil. Trans. R. Soc. A 376, 20170311 (2018) [arXiv:1801.01398].

\bibitem{CO2002} P. Grangier, ``Contextual objectivity: a realistic interpretation of quantum mechanics'', European Journal of Physics 23:3, 331 (2002) [arXiv:quant-ph/0012122]. 

 \end{references}
\end{document}